\documentclass[aps,prl,preprint,a4paper,showkeys,showpacs,superscriptaddress]{revtex4-1}
\usepackage{amssymb}
\usepackage{amsmath}
\usepackage{amsfonts}

\usepackage{graphicx}

\usepackage{braket}
\begin{document}

\newcommand{\sub}[1]{\ensuremath{_{\mathrm{#1}}}}
\newcommand{\sps}[1]{\ensuremath{^{\mathrm{#1}}}}

\title{Quantum Information Capsule and Information Delocalization by Entanglement in Multiple-qubit Systems}

\author{Koji Yamaguchi}
\affiliation{Graduate School of Science, Tohoku University,\\ Sendai, 980-8578, Japan}
\author{Naoki Watamura}
\affiliation{Department of Mathematics, Shanghai University,\\ 99 Shangda Road, Shanghai 200444, People's Republic of China}
\author{Masahiro Hotta}
\affiliation{Graduate School of Science, Tohoku University,\\ Sendai, 980-8578, Japan}

\begin{abstract}
Where do entangled multiple-qubit systems store information? For information injected into a qubit, this question is nontrivial and interesting since the entanglement delocalizes the information. 
So far, a common picture is that of a qubit and its purification partner sharing the information quantum mechanically. 
Here, we introduce a new picture of a single qubit in the correlation space, referred to as quantum information capsule (QIC), confining the information perfectly.
This picture is applicable for the entangled multiple-qubit system in an arbitrary state. 
Unlike the partner picture, 
in the QIC picture, by swapping the single-body state, leaving other subsystems untouched, the whole information can be retrieved out of the system. 
After the swapping process, no information remains in the system.
\end{abstract}
\maketitle

\textit{Introduction.}---
Quantum information storage plays a crucial role in modern physics. 
During the black hole evaporation process, information is lost in the semi-classical approximation \cite{BHIP}, while it should be preserved in unitary theories. 
Since the AdS/CFT correspondence \cite{adscft} suggests the unitarity of the process, it is important to investigate where information is stored. There is little consensus on the information storage and several candidates are proposed such as, the Hawking radiation itself \cite{Page_curve, AMPS}, hidden messengers in it \cite{BQML}, black hole quasi-normal modes \cite{Christian}, soft hairs \cite{softhair,HNY}, and the zero point fluctuation \cite{Wilczek} as the purification partner of the Hawking radiation \cite{HSU}. 
Information stored in quantum systems is generally scrambled by the time evolution. 
Black holes are conjectured to be fast scramblers \cite{SS}. 
As a measure of scrambling, tripartite information is often adopted \cite{TI}. It is closely related to out-of-order correlators, which are commonly used to diagnose quantum chaos \cite{MSS}. 
Quantum memory is also essential for quantum information technologies such as quantum computation \cite{NC}, quantum repeaters \cite{QR} for quantum network \cite{QI}, quantum cryptography \cite{BB84}, and quantum authentication \cite{QA}. Experimentally, storage and retrieval of information have been demonstrated in
a single atom \cite{HPS},
rare-earth ion doped solids \cite{PJ,MG},
nitrogen-vacancy centers in diamond \cite{CG,YK},
and vapor atoms \cite{DSD1,DSD2}.

\begin{figure}[h]
 \includegraphics[height=2cm]{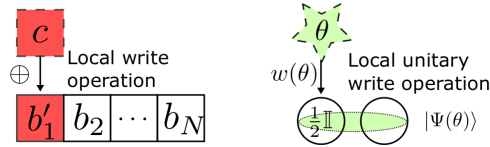}
\caption{\label{fig_cl}[Left] Localized information in classical memories. Classical information $c$ (red dashed square) is written in the first register (red solid square) as $b_1'=b_1\oplus c$. At the first register, the original information is recovered as $c=b_1\oplus b_1'$. [Right] Delocalization of information in the Bell state. Operating a local unitary $w(\theta)$ on the first qubit, information of a real parameter $\theta$ (green dashed star) is imprinted in a two-qubit system in the Bell state (solid circles). The reduced state of the first qubit is given by $\frac{\mathbb{I}}{2}$ and independent of $\theta$. The information is delocalized and hidden in nonlocal correlations (green dotted ellipse).}
\end{figure}

Despite such importance, it remains elusive where an entangled quantum system stores information. 
Let us first remind classical memories. 
In classical systems, information is localized. For example, suppose a device stores $N$-bit information. Its state is described by at least an $N$-length binary number $b=b_1b_2\cdots b_N$ where $b_n=0,1$. By using the exclusive disjunction $\oplus$, one can write a single-bit information $c$ on the first register. The state becomes $b'\equiv b_1' b_2\cdots b_N$ with $b_1'\equiv b_1\oplus c$. At the first register, $c$ is obtained by a local operation $c= b_1\oplus b_1'$, meaning that the information is stored locally. See the left picture in Fig.~\ref{fig_cl}. 
Now, let us consider a quantum memory described by an $N$-qubit system initially in an entangled state $\ket{\Psi}$. Information of a real unknown parameter $\theta$ is injected into the first qubit by a local unitary write operation $W(\theta)\equiv w(\theta)\otimes \mathbb{I}^{\otimes N-1}$, where $w(\theta)\equiv e^{-i\theta\sigma_z}$ and $\mathbb{I}$ denotes the identity operator for a qubit. Here and hereafter, $\{\sigma_i\}_{i=1}^3$ denotes the Pauli operators on a single qubit.
After the local write operation, the system evolves into $\ket{\Psi(\theta)}\equiv W(\theta)\ket{\Psi}$. 
The quantum Fisher information quantifies the precision for the estimation of the parameter from the state \cite{CWH}. 
For pure states, it is defined by $ F(\theta)\equiv 4(\braket{\partial_\theta\Psi(\theta)|\partial_\theta\Psi(\theta)}-\left|\braket{\Psi(\theta)|\partial_\theta\Psi(\theta)}\right|^2)$, which is independent of $\theta$ in this case, and given by $F=4(1-\braket{\Psi|\sigma_z \otimes \mathbb{I}^{\otimes N-1}|\Psi}^2)$. 
Unless $\Braket{\Psi|\sigma_z \otimes \mathbb{I}^{\otimes N-1}|\Psi}^2= 1$, information of $\theta$ is imprinted to the $N$-qubit system.
In analogy with classical memories, one might expect that it is possible to extract the information from the first qubit. 
In general, however, entanglement delocalizes the information \cite{AF}. 
As an example, let us consider a two-qubit system in a Bell state $\ket{\Psi}=\frac{1}{\sqrt{2}}\left(\ket{0}\ket{0}+\ket{1}\ket{1}\right)$, where $\ket{0}$ and $\ket{1}$ are $\sigma_z$'s eigenvectors, with eigenvalues $1$ and $-1$, respectively.
After the write operation, the system is in $\ket{\Psi(\theta)}=\frac{1}{\sqrt{2}}(e^{-i\theta}\ket{0}\ket{0}+e^{i\theta}\ket{1}\ket{1})$ and the Fisher information is nonzero: $F=4$.
One of the qubits has no information of $\theta$: the reduced state for the first qubit after the write operation is given by $\rho=\frac{\mathbb{I}}{2}$ and $\theta$-independent.
Thus, the information is stored in the nonlocal correlations of the two qubits. The information is delocalized as denoted in the right picture in Fig.~\ref{fig_cl}. It should be noted that this delocalization is different from the scrambling since no time evolution of the system has been taken into account so far.

For macroscopic quantum systems in pure states, it is difficult to avoid the delocalization of information,
since the smaller subsystem is almost maximally entangled with the other subsystem with high probability, as is proven in the famous theorem \cite{Lubkin,Lloyd_Pagels,Page}. Thus, the reduced state for the smaller subsystem is proportional to the identity operator in high precision and invariant under local write operations. The information is delocalized and shared by a macroscopic large number of qubits \cite{AS}. 

\begin{figure}[h]
 \includegraphics[height=3cm]{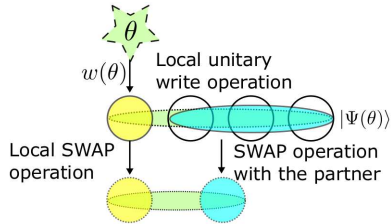}
\caption{\label{fig_partner} Information extraction using partners. The delocalized information of $\theta$ (green dotted ellipse) can be extracted by swapping the states of the first qubit (solid yellow circle) and its purification partner (solid blue ellipse) for the states of external two qubits (dotted yellow and blue circles).}
\end{figure}

Our aim is to extract the delocalized information without loss. 
There is a simple way by using the correlation space \cite{CS1,CS2}. 
The correlation space is a virtual state space defined by correlation functions of operators. 
Quantum operations on the virtual qubits are achieved by
operations on real qubits which affect the correlation functions. By using this property, measurement-based quantum computation has been developed.
By using the Schmidt decomposition for a given state $\ket{\Psi}$, it is always possible to find a two-dimensional sub-Hilbert space that purifies the real first qubit.
We refer to this two-dimensional sub-Hilbert space as the purification partner of the first qubit.
The composite system of the first qubit and its purification partner corresponds to a two-qubit system in the correlation space. 
After the local write operation $W(\theta)$ on the first real qubit, the two-qubit system in the correlation space remains pure. 
The whole information in the virtual qubit state is extracted by choosing an appropriate interaction between the $N$-qubit system and an external two-qubit to attain the quantum swap protocol \cite{NC}. In quantum mechanics, such a complete information extraction means that no information is left in the original system due to the no-cloning theorem \cite{no-cloning}. Fig.~\ref{fig_partner} shows this swapping protocol. 

\begin{figure}[h]
 \includegraphics[height=3cm]{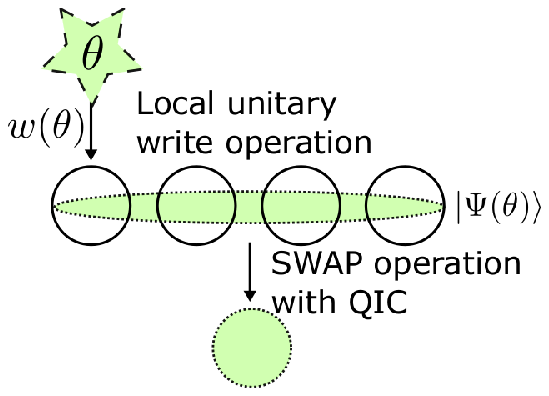}
\caption{\label{fig_qic} Information extraction using a QIC. The whole information of $\theta$ can be extracted by swapping the state of a QIC (dotted green ellipse) for the state of an external single qubit (dotted green circle).}
\end{figure}

In this Letter, we show that, for the $N$-qubit system in an arbitrary fixed state $\ket{\Psi}$, there exists a variety of partner pictures, which is
characterized by continuous parameters. 
Each pair contains the whole information of $\theta$ in a pure state, and shares different amount of entanglement depending on the parameters. By taking appropriate values of parameters, we show that the entanglement vanishes. This implies that a single virtual-qubit in the correlation space, referred to as quantum information capsule (QIC), confines the whole information in a pure state, as is depicted in Fig.~\ref{fig_qic}.
This QIC is a simple answer to the question of where information is stored. 
One might expect that, for an arbitrary state, maximal entangled partners could be obtained by taking other parameters.
Actually, it is not generally the case as opposed to QIC, as shown later. This fact makes the success of finding a QIC in any state nontrivial. It is consistent with prior research \cite{nohide,nomask} showing that quantum information can be hidden completely from subsystems only for specific states, though their setups are different from ours.

QIC has the following applications: (i) Reduction in extraction cost: We just need to prepare one external qubit in order to swap the state of a QIC, while two qubits are required externally in the case of partners.
(ii) Application to quantum authentication: Suppose an approver prepares a complicated state $\ket{\Psi}$ and asks an applicant to imprint $\theta$.
The approver will successfully obtain the information of $\theta$ by measuring QIC operators. The third party, who is ignorant of QIC which depends on $\ket{\Psi}$, will fail since the information is hidden in complicated nonlocal correlations. 
(iii) Application to quantum deep learning in entangled states: To deal with big data, macroscopic systems are required. QIC will be a unit of memory.
(iv) A new way to investigate nonlocal correlations: QIC tells us how nonlocal correlations are affected by a local disturbance. 
(v) Application to quantum chaos: Suppose the evolution of a system scrambles the information after the write operation. The increase in the complexity of QIC operators reflects the complexity of chaos. 
(vi) Application to black hole information loss problem: To analyze the black hole evaporation process, various qubit models have been proposed \cite{HNY, SDM,SBG,SGA}. It would be interesting to investigate how information is conserved and scrambled in those models by using the time evolution of QICs.

\textit{Existence of QIC.}---
In an $N$-qubit system, a single qubit in the correlation space is characterized by a set of traceless Hermitian operators $\{\Sigma_i\}_{i=1}^3$ satisfying
\begin{align}
 \Sigma_i\Sigma_j= \delta_{ij}\mathbb{I}^{\otimes N} +i\sum_{k=1}^3\epsilon_{ijk} \Sigma_k.\label{eq_cr_qubit}
\end{align} 
It should be noted that another set of operators $\{\Sigma_i'\}_{i=1}^3$ defined by $\Sigma_i'\equiv U^\dag \Sigma_i U$ with a unitary operator $U$ generated by $\{\Sigma_i\}_{i=1}^3$ represents the equivalent virtual-qubit. For example, $\{\sigma_i\otimes \mathbb{I}^{\otimes N-1}\}_{i=1}^3$ and $\{\sigma_x\otimes \mathbb{I}^{\otimes N-1}, -\sigma_z\otimes \mathbb{I}^{\otimes N-1}, \sigma_y\otimes \mathbb{I}^{\otimes N-1}\}$ are equivalent and have no physical difference. 
For a virtual qubit $A$ characterized by a set of traceless Hermitian operators $\{\Sigma_i^{(A)}\}_{i=1}^3$ satisfying Eq.~(\ref{eq_cr_qubit}), its partner qubit $B$ in a pure state $\ket{\Psi}$ is defined by a set of traceless Hermitian operators $\{\Sigma_i^{(B)}\}_{i=1}^3$ satisfying the following three conditions: (i) Algebra: it satisfies Eq.~(\ref{eq_cr_qubit}), (ii) Locality: $[\Sigma_j^{(A)},\Sigma_j^{(B)}]=0$ for $i,j=1,2,3$, (iii) Purification: the correlation space state defined by $\rho_{AB}\equiv \frac{1}{4}\sum_{\mu,\nu=0}^3\braket{\Sigma_\mu ^{(A)}\Sigma^{(B)}_\nu}\sigma_\mu\otimes\sigma_\nu$ is pure. Here we have defined $\Sigma_0^{(A)}\equiv\Sigma_0^{(B)}\equiv \mathbb{I}^{\otimes N}$, $\sigma_0\equiv\mathbb{I}$ and $\Braket{O}\equiv \braket{\Psi|O|\Psi}$ for an operator $O$.

Let us identify the partner of $\Sigma^{(A)}_i\equiv \sigma_i\otimes \mathbb{I}^{\otimes N-1}$. Consider the Schmidt decomposition of $\ket{\Psi}$ with respect to the tensor product structure $\mathcal{H}\otimes \mathcal{H}^{\otimes N-1}$, where $\mathcal{H}$ denotes a two-dimensional Hilbert space for a qubit. Its general form is given by $\ket{\Psi}=\sum_{i=1}^2\sqrt{p_i}\ket{\phi_i} \ket{\psi_i}$, where $\{\ket{\phi_i}\}_{i=1}^2$ and $\{\ket{\psi_i}\}_{i=1}^2$ are sets of orthonormal vectors, and $\{p_i\}_{i=1}^2$ is a probability distribution. In order to consider an entangled memory, we assume $p_i\neq 0$.  For entangled qubits, a single-qubit unitary operation affects the nonlocal correlations with the partner qubit. By using the vectors, the partner qubit in $\ket{\Psi}$ is constructed as $\Sigma^{(B)}_x\equiv \mathbb{I}\otimes(\ket{\psi_0}\bra{\psi_1}+\ket{\psi_1}\bra{\psi_0})$, $\Sigma^{(B)}_y\equiv \mathbb{I}\otimes(i\left(-\ket{\psi_0}\bra{\psi_1}+\ket{\psi_1}\bra{\psi_0}\right))$ and $\Sigma^{(B)}_z\equiv\mathbb{I}\otimes( \ket{\psi_0}\bra{\psi_0}-\ket{\psi_1}\bra{\psi_1})$. For the state $\ket{\Psi(\theta)}$, the sets of operators $\{\Sigma_i^{(A)}\}_{i=1}^3$ and $\{\Sigma_i^{(B)}\}_{i=1}^3$ are partners since $W(\theta)=e^{-i\theta\Sigma_z^{(A)}}$ corresponds to a local unitary operation on the first virtual qubit $A$. Thus, the whole information of $\theta$ is confined in this two-qubit system in the correlation space. By using a unitary operator $U$ generated by $\Sigma^{(A)}_z$ and $\{\Sigma^{(B)}_i\}_{i=1}^3$, we have another pair of partners $\{\Sigma^{(S')}_i\}_{i=1}^3$, where $\Sigma^{(S')}_i\equiv U^\dag \Sigma_i^{(S)}U$ for $S=A,B$. Since $\Sigma^{(A')}_z=U^\dag\Sigma^{(A)}_zU=\Sigma^{(A)}_z$ holds, the write operation $W(\theta)=e^{-i\theta\Sigma^{(A')}_z}$ corresponds to a local write operation on the virtual qubit $A'$ in the correlation space.  Each pair represents a different way of storing information with different amount of entanglement.

A QIC is a virtual qubit which confines the whole information in a pure state. 
To prove the existence of QIC, it is sufficient to find a unitary operator $U$ which commutes with $\Sigma^{(A)}_z$ and satisfies $U\ket{\Psi}=\ket{\phi}\ket{\psi}$, where $\ket{\phi}$ and $\ket{\psi}$ are pure states for a single qubit and $(N-1)$ qubits, respectively. For such a unitary operator, $A'$ corresponds to a QIC. The QIC state in the correlation space is given by
\begin{align}
& \rho_{A'}(\theta)=\frac{\sum_{\mu=0}^4 \braket{\Psi(\theta)|\Sigma^{(A')}_\mu|\Psi(\theta)}\sigma_\mu}{2}\nonumber\\
&=\frac{\sum_{\mu=0}^4 \braket{\phi|w(\theta)^\dag\sigma_\mu w(\theta)|\phi}\sigma_\mu}{2}=\ket{\phi(\theta)}\bra{\phi(\theta)}, 
\end{align}
where we have defined $\ket{\phi(\theta)}\equiv w(\theta)\ket{\phi}$. 
For proof of the existence of such a unitary operator, let us consider $U=e^{-ig\Sigma_z^{(A)}\tilde{\Sigma}^{(B)}_y}$, where $g$ is a real number. Here, we have introduced a new set of operators $\{\tilde{\Sigma}^{(B)}_i\}_{i=1}^3$ for the qubit $B$ as
\begin{equation}
 \begin{split}
  \tilde{\Sigma}_x^{(B)}&\equiv\mathbb{I}\otimes( \alpha_0\alpha_1^*\ket{\psi_0}\bra{\psi_i}+\alpha_0^*\alpha_1\ket{\psi_1}\bra{\psi_0}), \\
 \tilde{\Sigma}_y^{(B)}&\equiv \mathbb{I}\otimes(i(-\alpha_0\alpha_1^*\ket{\psi_0}\bra{\psi_1}+\alpha_0^*\alpha_1\ket{\psi_1}\bra{\psi_0})), \\
 \tilde{\Sigma}^{(B)}_z& \equiv \Sigma^{(B)}_z,
 \end{split}
\end{equation}
where 
\begin{align}
 \alpha_i\equiv
\begin{cases}
 \frac{\braket{0|\phi_i}}{|\braket{0|\phi_i}|} \, &(\text{if }\braket{0|\phi_i}\neq 0)\\
 1&(\text{otherwise}) 
\end{cases}
\end{align}
are complex numbers of unit modulus. By using them, let us define partners as $\Sigma_i^{(A')}\equiv U^\dag \sigma_i\otimes \mathbb{I}^{\otimes N-1}U$ and $\Sigma_i^{(B')}\equiv U^\dag \tilde{\Sigma}_i ^{(B)} U$.
To quantify the entanglement between the partners $A'B'$, let us calculate the purity of the virtual qubit $B'$ given by $\mathrm{Tr}(\rho_{B'}^2)=\frac{1}{2} \left(1+\sum_{i=1}^3\braket{\Sigma^{(B')}_i}^2\right)$. By using $U=\cos{g}\mathbb{I}^{\otimes N}-i\sin{g} \Sigma_z^{(A)}\tilde{\Sigma}_y^{(B)}$, each operator is calculated as
\begin{align}
 \Sigma_x^{(B')}&=\cos{2g}\tilde{\Sigma}_x^{(B)}+\sin{2g}\Sigma_z^{(A)}\tilde{\Sigma}_z^{(B)},\quad  \Sigma_y^{(B')}=\tilde{\Sigma}_y^{(B)}, \quad \Sigma_{z}^{(B')}= \cos{2g}\tilde{\Sigma}_z^{(B)}-\sin{2g}\Sigma_z^{(A)}\tilde{\Sigma}_x^{(B)}.
\end{align}
Their expectation values are given by
\begin{align}
 \Braket{\Sigma_x^{(B')}}&=\left(2\left|\braket{0|\phi_0}\right|^2-1\right)\sin{2g},\label{eq_sigmax}\\
 \Braket{\Sigma_y^{(B')}}&=0, \label{eq_sigmay}\\
 \Braket{\Sigma_z^{(B')}}&=\left(2p_0-1\right) \cos{2g} -4\sqrt{p_0(1-p_0)}\left|\Braket{0|\phi_0}\right|\sqrt{1-\left|\Braket{0|\phi_0}\right|^2}\sin{2g},\label{eq_sigmaz}
\end{align}
where we have used $p_0+p_1=1$, $\braket{\phi_0|\sigma_z|\phi_0}+\braket{\phi_1|\sigma_z|\phi_1}=\mathrm{Tr}(\sigma_z)=0$, $\sigma_z=2\ket{0}\bra{0}-\mathbb{I}$ and $\left|\Braket{0|\phi_0}\right|^2+\left|\Braket{0|\phi_1}\right|^2=\Braket{0|0}=1$. 
Combining equations (\ref{eq_sigmax}), (\ref{eq_sigmay}) and (\ref{eq_sigmaz}), we get $ \sum_{i=1}^3 \braket{\Sigma^{(B')}_i}^2= a+b\cos{4g}+c\sin{4g}$,
where
\begin{align}
 a &=\frac{1}{2}\left(1+(2p_0-1)^2(2\left|\Braket{0|\phi_0}\right|^2-1)^2\right) ,\\
 b &=(2p_0-1)^2-a,\\ 
 c&=-4(2p_0-1)\sqrt{p_0(1-p_0)} \left|\Braket{0|\phi_0}\right|\sqrt{1-\left|\Braket{0|\phi_0}\right|^2}.
\end{align}
Since $b^2+c^2=(1-a)^2$ holds, we get
\begin{align}
 \mathrm{Tr}\left(\rho_{B'}^2\right)&=\frac{1}{2}\left(1+a+(1-a)\cos{(4g-d)}\right),
\end{align}
where we have defined $d\equiv \mathrm{Arctan}{\frac{c}{b}}$.
 It implies the purity takes any value in $[a,1]$ with an appropriate $g$. Taking $g=\frac{d}{4}$, the virtual qubit $B'$ is in a pure state. Therefore, the virtual qubit $A'$ is a QIC. 

By using the family of pairs, it is also possible to show that maximally entangled partners exist if and only if $\braket{\Sigma_z^{(A)}}=0$. The partners are maximally entangled if and only if $\mathrm{Tr}(\rho_{A'}^2)=\mathrm{Tr}(\rho_{B'}^2)=\frac{1}{2}$, which is equivalent to $
\braket{\Sigma_i^{(A')}}=\braket{\Sigma_i^{(B')}}=0$ for $i=1,2,3$. Since $\braket{\Sigma_z^{(A)}}=\braket{\Sigma_z^{(A')}}$ holds, the maximally entangled partners exist only if $\braket{\Sigma_z^{(A)}}=0$. If this condition is satisfied, $a=\frac{1}{2}$ holds, meaning that $\mathrm{Tr}(\rho_{B'}^2)=\frac{1}{2}$ with $g=\frac{d+\pi}{4}$.

The information of $\theta$ confined in a QIC can be extracted by a swap operation. Let us first remind the swap operation in a two-qubit system, which is defined by $U_{\mathrm{swap}}=\sum_{i,j=0,1}\ket{i}\bra{j}\otimes \ket{j}\bra{i}$. For arbitrary states $\ket{\phi_1}$ and $\ket{\phi_2}$, this operation swaps one for another: $U_{\mathrm{swap}}\ket{\phi_1}\ket{\phi_2}=\ket{\phi_2}\ket{\phi_1}$. By using the Pauli operators, $U_{\mathrm{swap}}$ is described in another way: $U_{\mathrm{swap}}=\frac{1}{2}\sum_{\mu=0}^4 \sigma_\mu\otimes\sigma_\mu$. To swap the state of a QIC outside, swap operation is constructed as $U_{\mathrm{swap}}=\frac{1}{2}\sum_{\mu=0}^4\Sigma_\mu \otimes \sigma_\mu$, where we have defined $\Sigma_0\equiv \mathbb{I}^{\otimes N}$, $\{\Sigma_i\}_{i=1}^3$ is the operators characterizing the QIC, $\sigma_0\equiv \mathbb{I}$ and $\{\sigma_i\}_{i=1}^3$ are the Pauli operators for an external qubit system. 

\textit{Non-uniqueness of QIC.}--- 
By using the unitary operator $U$ such that $U\ket{\Psi}=\ket{\phi}\ket{\psi}$ and $[U,\Sigma_z^{(A)}]=0$, operators characterizing a QIC are constructed as $\Sigma_i\equiv U ^\dag\sigma_i\otimes \mathbb{I}^{\otimes N-1}U$.
It is always possible to construct another QIC which is inequivalent to the original one. 
Suppose an Hermitian operator $\mathcal{O}$ on $N-1$ qubits satisfies $\mathcal{O}\neq \mathbb{I}^{\otimes N-1}$, $\mathcal{O}^2=\mathbb{I}^{\otimes N-1}$ and $\mathcal{O}\ket{\psi}=\ket{\psi}$. For example, $\mathcal{O}=\ket{\psi}\bra{\psi}-\sum_{i=2}^{2^{N-1}}\ket{\psi^\perp_i}\bra{\psi^\perp_i}$ satisfies the requirements, where $\{\psi^\perp_i\}_{i=2}^{2^{N-1}}$ are orthonormal vectors orthogonal to $\ket{\psi}$.
A qubit in the correlation space characterized by operators $\Sigma'_x \equiv U^\dag \sigma_x \otimes \mathcal{O}U$, $\Sigma'_y \equiv U^\dag \sigma_y \otimes \mathcal{O} U$ and $\Sigma'_z\equiv U^\dag \sigma_z\otimes \mathbb{I}^{\otimes N-1}U$ is also a QIC, since $W(\theta)=e^{-i\theta\Sigma'_z}$ and $\braket{\Psi(\theta)|\Sigma'_i|\Psi(\theta)}=\braket{\phi(\theta)|\sigma_i|\phi(\theta)} $ holds for $i=1,2,3$. Note that they are traceless and satisfy $\Sigma'_i{}^{\dag }=\Sigma'_i$ and Eq.~(\ref{eq_cr_qubit}). 
To prove that the QIC defined by $\{\Sigma'_i\}_{i=1}^3$ is inequivalent to the QIC defined by $\{\Sigma_i\}_{i=1}^3$, let us consider a unitary operator $V$ generated by $\{\Sigma_i\}_{i=1}^3$. The set of operators $\{V^\dag\Sigma_i V\}_{i=1}^3$ also characterizes the QIC defined by $\{\Sigma_i\}_{i=1}^3$. For any $V$, $V^\dag \Sigma_i V=U{}^\dag\tilde{\sigma}_i\otimes \mathbb{I}^{\otimes N-1} U$ holds for some set of traceless Hermitian operators $\{\tilde{\sigma}_i\}_{i=1}^3$ satisfying $\tilde{\sigma}_i\tilde{\sigma}_j=\delta _{ij}\mathbb{I}+i\sum_{k=1}^3\epsilon_{ijk}\tilde{\sigma}_k$. Since $\mathcal{O}\neq \mathbb{I}^{\otimes N-1}$, the QICs characterized by $\{\Sigma_i\}_{i=1}^3$ and $\{\Sigma'_i\}_{i=1}^3$ are inequivalent with each other. 
The non-uniqueness of QIC shows that there are various ways to extract the information in pure states as we will explicitly see below.

\textit{Example: the GHZ state.}---
As an instructive example, let us consider QIC in the Greenberger--Horne--Zeilinger (GHZ) state for three qubits: $\ket{\text{GHZ}}=\frac{1}{\sqrt{2}}(\ket{+}\ket{+}\ket{+}+\ket{-}\ket{-}\ket{-})$ with $\ket{\pm}\equiv \frac{1}{\sqrt{2}}(\ket{0}\pm\ket{1})$. Since each qubit is maximally entangled with other qubits, the reduced state is invariant under any local write operation. Nevertheless, $\ket{\text{GHZ}(\theta)}\equiv W(\theta)\ket{\text{GHZ}}$ contains information of the unknown parameter $\theta$ since the Fisher information is $F=4$. A unitary operator $U\equiv \mathbb{I}\otimes \ket{+}\bra{+}\otimes \mathbb{I}+\sigma_z\otimes \ket{-}\bra{-}\otimes \mathbb{I}$ satisfies $U\ket{\text{GHZ}(\theta)}=w(\theta)\ket{+}\otimes \frac{1}{\sqrt{2}}(\ket{+}\ket{+}+\ket{-}\ket{-})$. Thus, $\{\Sigma_i\equiv U^\dag \sigma_i\otimes \mathbb{I}^{\otimes 2}U\}_{i=1}^3$ characterizes a QIC, which is given by
\begin{equation}
\begin{split}
 \Sigma^{(1)}_x&=\sigma_x\otimes \sigma_x\otimes \mathbb{I},\\
 \Sigma^{(1)}_y&=\sigma_y\otimes \sigma_x\otimes \mathbb{I}, \\
\Sigma^{(1)}_z&=\sigma_z\otimes \mathbb{I}\otimes \mathbb{I}.
\end{split}\label{eq_ghz_qic1}
\end{equation}

Now, consider an Hermitian operator $\mathcal{O}=\sigma_x\otimes \sigma_x$ satisfying the requirements imposed in the previous section. Thus, the operators
\begin{equation}
\begin{split}
\Sigma^{(2)}_x &\equiv U^\dag \sigma_x\otimes \mathcal{O}U=\sigma_x\otimes \mathbb{I}\otimes \sigma_x,\\
 \Sigma^{(2)}_y &\equiv U^\dag \sigma_y\otimes \mathcal{O}U=\sigma_y\otimes \mathbb{I}\otimes \sigma_x,\\
 \Sigma^{(2)}_z &\equiv U^\dag \sigma_z\otimes \mathbb{I}\otimes \mathbb{I}U=\sigma_z\otimes\mathbb{I}\otimes \mathbb{I}
\end{split}
\label{eq_ghz_qic2}
\end{equation}
characterize a different QIC. In this case, it is easy to see the QICs defined by Eqs.~(\ref{eq_ghz_qic1}) and (\ref{eq_ghz_qic2}) are related via an interchange of the second and third qubit, under which $\ket{\text{GHZ}(\theta)}$ is invariant. 

In order to extract the information of $\theta$, let us consider a swap unitary operation of QIC with an external qubit. If one wants to use the $i$th QIC, it is given by $ U^{(i)}\equiv \frac{1}{2}\sum_{\mu=0}^4\Sigma^{(i)}_\mu\otimes \sigma_\mu$ for $i=1,2$. Assuming the initial state of readout qubit is $\ket{\chi_0}$, $U^{(i)}\ket{\text{GHZ}(\theta)}\otimes \ket{\chi_0}=\ket{\chi_0}\frac{1}{\sqrt{2}}(\ket{+}\ket{+}+\ket{-}\ket{-})\otimes w(\theta)\ket{+}$ holds for $i=1,2$ after the swap operation, even though $U^{(1)}\neq U^{(2)}$. It implies that we can extract the information if we are accessible to either first and second qubits, or first and third qubits. This example shows that a different QIC gives a different way to process information.

\textit{Dynamics of QIC.}---
Let $H$ be the Hamiltonian of an $N$-qubit system. Assume that the system is initially in a pure state $\ket{\Psi}$, and the write operation $W(\theta)$ is performed instantaneously at $t=0$. Then, at $t>0$, the system evolves into $e^{-iHt}\ket{\Psi(\theta)}$. Information of $\theta$ imprinted in the $N$-qubit system is scrambled but conserved since the Fisher information is invariant under any unitary evolution. One may be worried that it would be difficult to identify how the system retains information. However, QIC provides a simple way to track scrambled information. Let $\{\Sigma_i\}_{i=1}^3$ be the associated operators of a QIC for $\ket{\Psi(\theta)}$. By using the QIC at $t=0$, a QIC at $t>0$ is given by $\{\Sigma_i(t)\}_{i=1}^3$, where $\Sigma_i(t)\equiv e^{-iHt}\Sigma_i e^{iHt}$. It should be noted that this is the time-reversed evolution of operators in the Heisenberg picture. This construction is applicable to any dynamics, and provides a direct way to identify a qubit in the correlation space that carries the locally injected information. 

\textit{Conclusions.}---
In this Letter, we have investigated the way to extract the delocalized information due to entanglement. Introducing virtual qubits in the correlation space, it is possible to adopt different pictures where various pairs of entangled partners share the information. We have shown that a virtual qubit, which we call a QIC, contains the whole information injected by a local unitary operation $w(\theta)\otimes\mathbb{I}^{\otimes N-1}$. A way to construct a QIC for an arbitrary state has been presented, which enables us to retrieve the delocalized information by a simple swap operation. 
Since QIC is non-unique, there is no definite place where the information is stored. 
In addition, the time evolution of QIC derived here directly shows how information is scrambled due to the evolution of system. All the results are straightforwardly extended to a write operation $w_{\vec{n}}(\theta)\otimes \mathbb{I}^{\otimes N-1}$, where $w_{\vec{n}}(\theta)\equiv e^{-i\theta \sum_{i=1}^3 n_i\sigma_i}$ with a real unit vector $\vec{n}=(n_1,n_2,n_3)$.

The authors thank Ursula Carow-Watamura and Takeshi Tomitsuka for discussions. This research was partially supported by JSPS KAKENHI Grant Numbers JP16K05311 (M.H.) and JP18J20057 (K.Y.), and by Graduate Program on Physics for the Universe of Tohoku University (K.Y.).

\end{document}